\documentstyle[12pt]{article}  
\begin{document}  \bibliographystyle{unsrt} \vbox{\vspace{10mm}}

\centerline{\Large \bf Eugene Wigner and Translational Symmetries}
\vspace{6mm}
\centerline{ Y.S.Kim\footnote{Internet: kim@umdhep.umd.edu}}
\centerline{\it Department of Physics, University of Maryland,}
\centerline{\it College Park, Maryland 20742, U.S.A.}
\vspace{6mm}

\begin{abstract}
As Einstein's $E = mc^{2}$  unifies the energy-momentum relation for
massive and massless particles, Wigner's little group unifies their
internal space-time symmetries. It is pointed out that translational
symmetries play essential roles both in formulating the problem,
and in deriving the conclusions. The input translational symmetry
is the space-time translation, and the output translational symmetry
is built in the $E(2)$-like little group for massless particles.
The translational degrees of freedom in this little group are gauge
degrees of freedom.  It should be noted that a number of condensed
matter physicists played important supporting and supplementary
roles in Wigner's unification of internal space-time symmetries of
massive and massless particles.
\end{abstract}

\section{Introduction}\label{intro}
Eugene Wigner wrote many fundamental papers.  However, if I am forced to
name the most important paper he wrote, I have to mention his 1939 paper
on representations of the inhomogeneous Lorentz group which is often
called the Poincar\'e group~\cite{wig39}.  In this paper, he introduces
the ``little group'' which governs the internal space-time symmetry of
relativistic particles.  The physical implication of the little groups
was not fully recognized when Wigner received the 1963 Nobel Prize
in Physics.  The best way to see the scientific value of Wigner's
little group is to compare it with Einstein's work using the following
table.

\vspace{4mm}
\begin{center}
\begin{tabular}{ccc}
Massive, Slow & COVARIANCE & Massless, Fast \\[2mm]\hline
{}&{}&{}\\
$E = p^{2}/2m$ & Einstein's
$E = \sqrt{m^{2} + p^{2}}$
& $E = cp$ \\[4mm]\hline
{}&{}&{}  \\
$S_{3}$ & {}  &    $S_{3}$ \\ [-1mm]
{} & Wigner's Little Group & {} \\[-1mm]
$S_{1}, S_{2}$ & {} & Gauge Trans. \\[4mm]\hline
\end{tabular}
\end{center}
\vspace{4mm}

\noindent
We shall call the above table {\em Einstein-Wigner Table}.  While
Einstein's $E = mc^{2}$ unifies the energy-momentum relations
for massive and massless particles, Wigner's little group unifies the
internal space-time symmetries of massive and massless particles.
A massive particle has three rotational degrees of freedom, and they
are known as the spin degrees of freedom.  A massless particle has one
helicity degree of freedom.  Unlike massive particles, massless
particles have gauge degrees of freedom.  This table was first published
in 1986~\cite{hks86jm}, and I am the one who showed the table to Professor
Wigner in 1985 before it was published. In his 1939 paper~\cite{wig39},
Wigner noted that the little groups for massive and massless particles
are locally isomorphic to the three-dimensional rotation group and the
two-dimensional Euclidean groups respectively.  We shall
call them the $O(3)$-like little group for massive particles and
$E(2)$-like little group for massless particles.

Of course, the group
of Lorentz transformations plays the central role
in the little group formalism.  But we need more.  The purpose of
the present report is to examine carefully the role of translational
symmetries in completing the above table.  The translational symmetry is
very close to our daily life.
When we walk, we perform translations on our body.  Mathematically
speaking, however, translations are very cumbersome operations.
Indeed, there are many mathematical words associated with translations.
To name a new, we use the words affine groups, inhomogeneous
transformations, noncompact groups, semi-direct products, semi-simple
Lie groups, invariant subgroup, induced representations, solvable
groups, and more.
Thus, it is not unreasonable to expect some non-trivial physical
conclusions derivable from the translational symmetries.

In order to illustrate the mathematical complication from translational
degrees of freedom, let us construct a rotation matrix applicable to a
column vector $(x, y, 1)$:
\begin{equation}\label{rot}
R(\theta) = \pmatrix{\cos\theta & -\sin\theta & 0 \cr
\sin\theta & \cos\theta & 0 \cr 0 & 0 & 1} .
\end{equation}
Let us then consider the translation matrix:
\begin{equation}
T(a, b) = \pmatrix{1 & 0 & a \cr 0 & 1 & b \cr 0 & 0 & 1} .
\end{equation}
If we take the product $T(a, b) R(\theta)$,
\begin{equation}\label{eucl}
E(a, b, \theta) = T(a, b) R(\theta) =
\pmatrix{\cos\theta & -\sin\theta & a \cr
\sin\theta & \cos\theta & b \cr 0 & 0 & 1} .
\end{equation}
This is the Euclidean transformation matrix applicable to the
two-dimensional $x y$ plane.  The matrices $R(\theta)$ and $T(a,b)$
represent the rotation and translation subgroups respectively.  The
above expression is not a direct product because $R(\theta)$ does not
commute with $T(a,b)$.  The translations constitute an Abelian invariant
subgroup because two different $T$ matrices commute with each other,
and because
\begin{equation}
R(\theta) T(a,b) R^{-1}(\theta) = T(a',b') .
\end{equation}
The rotation subgroup is not invariant because the conjugation
$$T(a,b) R(\theta) T^{-1}(a,b)$$
does not lead to another rotation.

We can write the above transformation matrix in terms of generators.
The rotation is generated by
\begin{equation}\label{j3}
J_{3} = \pmatrix{0 & -i & 0 \cr i & 0 & 0 \cr 0 & 0 & 0} .
\end{equation}
The translations are generated by
\begin{equation}
P_{1} = \pmatrix{0 & 0 & i \cr 0 & 0 & 0 \cr 0 & 0 & 0} , \qquad
P_{2} = \pmatrix{0 & 0 & 0 \cr 0 & 0 & i \cr 0 & 0 & 0} .
\end{equation}
These generators satisfy the commutation relations:
\begin{equation}\label{e2com}
[P_{1}, P_{2}] = 0 , \qquad [J_{3}, P_{1}] = iP_{2},
\qquad [J_{3}, P_{2}] = -iP_{1} .
\end{equation}
This $E(2)$ group is not only convenient for illustrating the groups
containing an Abelian invariant subgroup, but also occupies an
important place in constructing representations for the little
group for massless particles, since the little group for massless
particles is locally isomorphic to the above $E(2)$ group.

In Sec. \ref{review}, we give a historical review of Wigner's little
groups. In Sec. \ref{contrac}, we shall see how the little group for
a massive particle can becomes a little group for massless particles
in the infinite-momentum/zero-mass limit.  In Sec. \ref{gauge}, we
review how the gauge degree of freedom of a massless particle is
associated with the translation-like degrees of freedom in the
$E(2)$-like little group for a massless particle.

\section{Historical Review of Wigner's Little Groups}\label{review}
Wigner was the first one to introduce the rotation group to
the quantum mechanics of atomic spectra~\cite{wig31}.  Since he had
a strong background in chemistry~\cite{szanton92}, he became also
interested in condensed matter physics.  He had a graduate student
named Frederick Seitz at Princeton University.  Wigner and Seitz
together published a paper on the constitution of metallic sodium
in 1933~\cite{seitz33}.

Seitz then started developing his own research line on space groups
applicable to solid crystals.  In his 1936 paper~\cite{seitz36},
Seitz discussed the augmentation of translation degrees of freedom
to the three-dimensional rotation group, just as we did for the
two-dimensional plane in Sec. \ref{intro} of the present paper.  The
results Seitz obtained were new at that time.  This seminal paper
opened up an entirely new research line on symmetries in crystals.  I
am not competent enough to write a review article on this research line,
but I can say that the physicists belonging to this genealogy later
made a decisive contribution to the physics of Wigner's little groups.
The purpose of this report is to discuss what role the condensed matter
physicists played in completing the table given in Sec. \ref{intro}.

In his 1939 paper~\cite{wig39}, Wigner replaced the three-dimensional
rotation group Seitz's paper by the four-dimensional Lorentz group.
By bringing in the translational degrees of freedom to the group of
Lorentz transformations, Wigner observed the importance of the
four-momentum in constructing representations.  He then focused his
attention to the subgroups of the Lorentz group whose transformations
leave the four-momentum of a given particle invariant.  The maximal
subgroup which leaves the four-momentum invariant is called the little
group.  The point is that a relativistic particle has its
Lorentz-covariant internal space-time structure, in addition to its
four-momentum.

The group of Lorentz transformations consists of three boosts and
three rotations.  The rotations therefore constitute a subgroup of
the Lorentz group.  If a massive particle is at rest, its four-momentum
is invariant under rotations.  Thus the little group for a massive
particle at rest is the three-dimensional rotation group.  Then what is
affected by the rotation?  The answer to this question is very simple.
The particle in general has its spin.  The spin orientation is going
to be affected by the rotation!

The rest-particle can now be boosted, and it will pick up non-zero
space-like momentum components.  The generators of the $O(3)$ little
group will also be boosted.  The boost will take the form of
conjugation by the boost operator.  This boost will not change the
Lie algebra of the rotation group and the boosted little group
will still leave the boosted four-momentum invariant.  We call this
the $O(3)$-like little group.

It is not possible to bring a massless particle to its rest frame.
In his 1939 paper~\cite{wig39}, Wigner observed that the little group
for a massless particle along the $z$ axis is generated by the
rotation generator around the $z$ axis, and two other generators.
If we use the four-vector coordinate $(x, y, z, t)$, the rotation
around the $z$ axis is generated by
\begin{equation}
J_{3} = \pmatrix{0 & -i & 0 & 0 \cr i & 0 & 0 & 0
\cr 0 & 0 & 0 & 0 \cr 0 & 0 & 0 & 0} ,
\end{equation}
and the other two generators are
\begin{equation}\label{n1n2}
N_{1} = \pmatrix{0 & 0 & -i & i \cr 0 & 0 & 0 & 0
\cr i & 0 & 0 & 0 \cr i & 0 & 0 & 0} ,  \quad
N_{2} = \pmatrix{0 & 0 & 0 & 0 \cr 0 & 0 & -i & i
\cr 0 & i & 0 & 0 \cr 0 & i & 0 & 0} .
\end{equation}
If we use $K_{i}$ for the boost generator along the i-th axis, these
matrices can be written as
\begin{equation}
N_{1} = K_{1} - J_{2} , \qquad N_{2} = K_{2} + J_{1} .
\end{equation}
Very clearly, the generators $J_{3}, N_{1}$ and $N_{2}$ generate
Lorentz transformations (boosts and rotations), and they satisfy the
commutation relations:
\begin{equation}\label{e2lcom}
[N_{1}, N_{2}] = 0 , \qquad [J_{3}, N_{1}] = iN_{2},
\qquad [J_{3}, N_{2}] = -iN_{1} .
\end{equation}
If we replace $N_{1}$ and $N_{2}$ by $P_{1}$ and $P_{2}$, then the
above set of commutation relations becomes the set given for the $E(2)$
group given in Eq.(\ref{e2com}).  This is the reason why the little
group for massless particles is $E(2)$-like.

It is not difficult to associate the rotation generator $J_{3}$ with
the helicity degree of freedom of the massless particle.   Then what
physical variable is associated with the $N_{1}$ and $N_{2}$ generators?
Wigner left this problem as a homework problem for younger generations.
Before attempting to solve this problem, let us note that there are at
least two more homework problems contained in Wigner's paper~\cite{wig39}.

\begin{itemize}

\item First, it is possible to interpret the Dirac equation in terms of
Wigner's representation theory~\cite{barg48}.  Then, why is it not
possible to find a place for Maxwell's equations in the same theory?

\item Second, as is shown by Inonu and Wigner~\cite{inonu53}, the
rotation group $O(3)$ can be contracted to $E(2)$.  Does this mean that
the $O(3)$-like little group can become the $E(2)$-like little group in
a certain limit?

\end{itemize}

We shall come back to the main question in Sec. \ref{gauge}, and we
shall deal with the second question Sec. \ref{contrac}.  Since the
physics of $N_{1}$, and $N_{2}$ was not known, there had been a tendency
in the past to construct representations for massless particles without
these degrees of freedom.  Indeed, in 1964~\cite{wein64},
Weinberg found a place for the electromagnetic tensor in Wigner's
representation theory.  He accomplished this by constructing from the
$SL(2,c)$ spinors all the representations of massless fields which are
invariant under the translation-like transformations of the E(2)-like
little group.  Weinberg stated in 1964 that the $N$-invariant state
vectors are gauge-invariant states, indicating that the $N$ operators
generate gauge transformations.  Weinberg did not elaborate on this
point in his 1964 papers~\cite{wein64}. We shall return to the question
of the $N_{1}$ and $N_{2}$ in Sec. \ref{gauge}.

\section{Contraction of O(3)-like Little Group to E(2)-like Little
Group}\label{contrac}
The contraction of $O(3)$ to $E(2)$ is well known and is often called
the Inonu-Wigner contraction~\cite{inonu53}.  The question is whether
the $E(2)$-like little group can be obtained from the $O(3)$-like
little group.  In order to answer this question, let us closely look
at the original form of the Inonu-Wigner contraction.  We start with
the generators of $O(3)$.  The $J_{3}$ matrix is given in Eq.(\ref{j3}),
and
\begin{equation}\label{o3gen}
J_{2} = \pmatrix{0&0&i\cr0&0&0\cr-i&0&0} , \qquad
J_{3} = \pmatrix{0&-i&0\cr i &0&0\cr0&0&0} .
\end{equation}
The Euclidean group $E(2)$ is generated by $J_{3}, P_{1}$ and $P_{2}$,
and their Lie algebra has been discussed in Sec.(\ref{intro}).

Let us transpose the Lie algebra of the $E(2)$ group.  Then $P_{1}$ and
$P_{2}$ become $Q_{1}$ and $Q_{2}$ respectively, where
\begin{equation}
Q_{1} = \pmatrix{0&0&0\cr0&0&0\cr i &0&0} , \qquad
Q_{2} = \pmatrix{0&0&0\cr0&0&0\cr0&i&0} .
\end{equation}
Together with $J_{3}$, these generators satisfy the
same set of commutation relations as that for
$J_{3}, P_{1}$, and $P_{2}$ given in Eq.(\ref{e2com})
\begin{equation}
[Q_{1}, Q_{2}] = 0 , \qquad [J_{3}, Q_{1}] = iQ_{2} , \qquad
[J_{3}, Q_{2}] = -iQ_{1} .
\end{equation}
These matrices generate transformations of a point on a circular
cylinder.  Rotations around the cylindrical axis are generated by
$J_{3}$.  The matrices $Q_{1}$ and $Q_{2}$ generate translations along
the direction of $z$ axis.  The group generated by these three matrices
is called the {\it cylindrical group}~\cite{kiwi87jm,kiwi90jm,misra76}.

We can achieve the contractions to the Euclidean and cylindrical groups
by taking the large-radius limits of
\begin{equation}\label{inonucont}
P_{1} = {1\over R} B^{-1} J_{2} B ,
\qquad P_{2} = -{1\over R} B^{-1} J_{1} B ,
\end{equation}
and
\begin{equation}
Q_{1} = -{1\over R}B J_{2}B^{-1} , \qquad
Q_{2} = {1\over R} B J_{1} B^{-1} ,
\end{equation}
where
\begin{equation}\label{bmatrix}
B(R) = \pmatrix{1&0&0\cr0&1&0\cr0&0&R}  .
\end{equation}
The vector spaces to which the above generators are applicable are
$(x, y, z/R)$ and $(x, y, Rz)$ for the Euclidean and cylindrical groups
respectively.  They can be regarded as the north-pole and equatorial-belt
approximations of the spherical surface respectively~\cite{kiwi87jm}.

Since $P_{1} (P_{2})$ commutes with $Q_{2} (Q_{1})$, we can consider the
following combination of generators.
\begin{equation}
F_{1} = P_{1} + Q_{1} , \qquad F_{2} = P_{2} + Q_{2} .
\end{equation}
Then these operators also satisfy the commutation relations:
\begin{equation}\label{commuf}
[F_{1}, F_{2}] = 0 , \qquad [J_{3}, F_{1}] = iF_{2} , \qquad
[J_{3}, F_{2}] = -iF_{1} .
\end{equation}
However, we cannot make this addition using the three-by-three matrices
for $P_{i}$ and $Q_{i}$ to construct three-by-three matrices for $F_{1}$
and $F_{2}$, because the vector spaces are different for the $P_{i}$ and
$Q_{i}$ representations.  We can accommodate this difference by creating
two different $z$ coordinates, one with a contracted $z$ and the other
with an expanded $z$, namely $(x, y, Rz, z/R)$.  Then the generators
become
\begin{equation}
P_{1} = \pmatrix{0&0&0&i\cr0&0&0&0\cr0&0&0&0\cr0&0&0&0} , \qquad
P_{2} = \pmatrix{0&0&0&0\cr0&0&0&i\cr0&0&0&0\cr0&0&0&0} .
\end{equation}
\begin{equation}
Q_{1} = \pmatrix{0&0&0&0\cr0&0&0&0\cr i &0&0&0\cr0&0&0&0} , \qquad
Q_{2} = \pmatrix{0&0&0&0\cr0&0&0&0\cr0&i&0&0\cr0&0&0&0} .
\end{equation}
Then $F_{1}$ and $F_{2}$ will take the form:
\begin{equation}\label{f1f2}
F_{1} = \pmatrix{0&0&0&i\cr0&0&0&0\cr i &0&0&0\cr0&0&0&0} , \qquad
F_{2} = \pmatrix{0&0&0&0\cr0&0&0&i\cr0&i&0&0\cr0&0&0&0} .
\end{equation}
The rotation generator $J_{3}$ takes the form
\begin{equation}\label{expl3}
J_{3} = \pmatrix{0&-i&0&0\cr i&0&0&0\cr0&0&0&0\cr0&0&0&0} .
\end{equation}
These four-by-four matrices satisfy the E(2)-like commutation relations
of Eq.(\ref{commuf}).

Now the $B$ matrix of Eq.(\ref{bmatrix}), can be expanded to
\begin{equation}\label{bmatrix2}
B(R) = \pmatrix{1&0&0&0\cr0&1&0&0\cr0&0&R&0\cr0&0&0&1/R}  .
\end{equation}
If we make a similarity transformation on the above form using the matrix
\begin{equation}\label{simil}
\pmatrix{1&0&0&0\cr0&1&0&0\cr0&0&1/\sqrt{2} &-1/\sqrt{2}
\cr0&0&1/\sqrt{2}&1/\sqrt{2}}  ,
\end{equation}
which performs a 45-degree rotation of the third and fourth coordinates,
then this matrix becomes
\begin{equation}\label{simil2}
\pmatrix{1&0&0&0\cr0&1&0&0\cr0&0 & \cosh\eta & \sinh\eta
\cr0 & 0 & \sinh\eta & \cosh\eta}  ,
\end{equation}
with $R = e^\eta$.  This form is the Lorentz boost matrix along the $z$
direction.  If we start with the set of expanded rotation generators
$J_{3}$ of Eq.(\ref{expl3}), and
\begin{equation}
J_{1} = \pmatrix{0&0&0&0\cr0&0&-i&0\cr0&i&0&0\cr0&0&0&0} , \quad
J_{2} = \pmatrix{0&0&i&0\cr0&0&0&0\cr-i&0&0&0\cr0&0&0&0} ,
\end{equation}
perform the same operation as the original Inonu-Wigner contraction
given in Eq.(\ref{inonucont}), the result is
\begin{equation}
N_{1} = {1\over R} B^{-1} J_{2} B ,
\qquad N_{2} = -{1\over R} B^{-1} J_{1} B ,
\end{equation}
where $N_{1}$ and $N_{2}$ are given in Eq.(\ref{n1n2}).

We are ultimately interested in giving the physical interpretation to
the Wigner row in the Einstein-Wigner table given in Sec. \ref{intro}.
In the meantime, it is clear now from this Section that $N_{1}$ and
$N_{2}$ are Lorentz-boosted rotation generators $J_{2}$ and $J_{1}$
respectively.  All we have to do next is to give a physical
interpretation to these operators.

\section{Translations and Gauge Transformations}\label{gauge}
As was noted in Sec. \ref{review}, it is possible to get the hint
that the $N$ operators generate gauge transformations from Weinberg's
1964 papers~\cite{wein64,hks82}.  But it was not until 1971 when
Janner and Janssen explicitly demonstrated that they generate gauge
transformations~\cite{janner71,kuper76}.  In order to fully appreciate
their work, let us compute the transformation matrix generated by
$N_{1}$, which takes the form
\begin{equation}\label{trans}
 \exp{(-iuN_{1})} = \pmatrix{1 & 0 &-u & u \cr 0 & 1 & 0 & 0 \cr
u & 0 & 1 - u^{2}/2 & u^{2}/2 \cr u & 0 & -u^{2}/2 & 1 + u^{2}/2} .
\end{equation}
If we apply this matrix to the four-vector to the four-momentum vector
\begin{equation}\label{4mom}
p = (0, 0, \omega, \omega)
\end{equation}
of a massless particle, the momentum remains invariant.  It therefore
satisfies the condition for the little group.  If we apply this matrix
to the electromagnetic four-potential
\begin{equation}
A = (A_{1}, A_{2}, A_{3}, A_{0}) \exp{(i(kz -\omega t))} ,
\end{equation}
the result is a gauge transformation.  This is what Janner and Janssen
discovered in their 1971 and 1972 papers~\cite{janner71}.

In order to see this without going through the generators, let us
rotate the four-momentum of Eq.(\ref{4mom}) around the $y$ axis using
the matrix
\begin{equation}\label{eq4}
R(\theta) = \pmatrix{\cos\theta & 0 & \sin\theta & 0 \cr
0 & 1 & 0 & 0 \cr
-\sin\theta & 0 & \cos\theta & 0 \cr 0 & 0 & 0 & 1} .
\end{equation}
We can come back to the original four-momentum by applying the inverse
$R(\theta)$, but the result will be trivial.  However, we can also
come back to the original momentum by a boost.  The boost matrix
needed for this process is
\begin{equation}\label{eq5}
\pmatrix{1 + 2[\sin(\theta/2)]^{2} & 0 &
-[\sin(\theta/2)]^{2}\tan(\theta/2) & -2 \tan(\theta/2) \cr
0 & 1 & 0 & 0 \cr
- [\sin(\theta/2)]^{2}\tan(\theta/2) & 0 &
1 + 2 [\tan(\theta/2)\sin(\theta/2)]^{2} &
2 [\tan(\theta/2)]^{2} \cr
- 2 \tan(\theta/2) & 0 &
- 2 [\tan(\theta/2)]^{2} & 1 + 2 [\tan(\theta/2)]^{2}} .
\end{equation}
The rotation $R(\theta)$ followed by the above boost
matrix leaves the four-momentum $p$ of Eq.(\ref{4mom}) invariant.
The resulting transformation matrix is
\begin{equation}
D(u) = \pmatrix{1 & 0 & \tan(\theta/2) & -\tan(\theta/2) \cr
0 & 1 & 0 & 0 \cr
- \tan(\theta/2) & 0 & 1 -
[\tan(\theta/2)]^{2}/2 & [\tan(\theta/2)]^{2}/2
\cr - \tan(\theta/2) & 0 & -[\tan(\theta/2)]^{2}/2 &
1 + [\tan(\theta/2)]^{2}/2} .
\end{equation}
If we replace $\tan(\theta/2)$ by $-u$, then the result will be
the transformation matrix given in Eq.(\ref{trans}) generated by
$N_{1}$.  We can carry out a similar calculation with $N_{2}$, and
the conclusion will be the same.

Indeed, the $N$ matrices generate gauge transformations.  If we
combine this with the result of Sec. \ref{contrac}, the conclusion is
that the Lorentz-boosted $J_{1}$ and $J_{2}$ become the generators of
gauge transformations in the limit of infinite momentum and/or
zero-mass~\cite{hks83}.  The operator $J_{3}$ remains invariant and
keeps serving as the helicity generator.  These results lead to the
Einstein-Wigner table given in Sec. \ref{intro}.

\section*{Acknowledgments}
I would like thank the organizers of this conference, particularly
Professor Tadeusz Lulek and Professor Wojciech Florek, for inviting me to
present this paper.  The Einstein-Wigner table given in Sec. \ref{intro}
is important to all physicists, but more so to particle theorists, and
I am one of them.  For this reason, it is a gratifying experience for
me to recognize the contributions made by condensed matter physicists
toward the construction of the table.  In 1987, I had a telephone
conversation with Professor Frederick Seitz.  He encouraged me to
develop a conference series which these days is called the
International Wigner Symposium.  The fifth meeting of this Symposium
will be held in Vienna in 1997.

I met Professor Aloysio Janner for the first time in 1978 during the
Seventh International Colloquium on Group Theoretical Methods in
Physics held in Austin, Texas.  Since then he has given me many
valuable advices a senior group theoretician.  However, he was too
modest to tell me about his own papers on Wigner's little group.  He
was still modest when I told him last month (July 1996) in Germany
about my plan to present this paper at this conference.  His modesty
made my job of finding his contribution very difficult, and I did not
know about his papers on this subject until I wrote with M. E. Noz on
the Poincar\'e group which was published in 1986~\cite{knp86}, even
though we managed to quote one of his papers in the book.  It is
indeed my personal pleasure to emphasize at this conference his role
in completing the Einstein-Wigner table given in Sec. \ref{intro}.
Professor Janner of course has a brilliant publication record in the
field of application of group theory to crystal physics.  I hope to
learn more from his papers in the future.

I am not known as a condensed matter theorist, but I am here because
I published seven papers with Eugene Wigner, and many people ask me
how I did it.  I met Professor Wigner while I was a graduate student
at Princeton University from 1958 to 1961.  I stayed there for one
more year as a post-doctoral fellow before joining the faculty of the
University of Maryland in 1962.  My advisor at Princeton was Sam Treiman,
and I wrote my PhD thesis on dispersion relations.  However, I did
my extra-curricular activity on Wigner's papers, particularly on his
1939 paper on representations of the Poincar\'e group~\cite{wig39}.  It
is not uncommon for one's extra-curricular activity to become his/her
life-time job.  Indeed, by 1985, I had completed the manuscript for
the above-mentioned book entitled {\em Theory and Applications of the
Poincar\'e Group}~\cite{knp86} with Marilyn Noz who has been my
closest colleague since 1970.

After writing this book in 1985, I approached Wigner again and asked
him whether I could start working on edited volumes of all the papers
he had written, but he had a better idea.  Wigner told me that he was
interested in writing new papers and that he had been looking for a
younger person who could collaborate with him.  He was interested in
many interesting problems.  However, I was only able to assist him on
two subjects.  One was on group contractions.  Wigner was particularly
eager to establish a connection between the Inonu-Wigner contraction
and his work on the Poincar\'e group.  He was also interested in
constructing Wigner functions which can be Lorentz-transformed.  I
was well prepared for the first problem, but I did not have a strong
background in the Wigner function.  On the other hand, I had a
formalism of harmonic oscillators which can be Lorentz-transformed,
and this formalism forms the back-bone of my book with Noz~\cite{knp86}.
It is not difficult to translate the oscillator formalism into the
language of Wigner functions. This was how I was able to write papers
with him.

During the period 1986 - 1991, I went to Princeton regularly to work
with him.  It takes three hours by train to go to Princeton from the
University of Maryland.  After this period, I became much smarter. I am
eternally grateful to Professor Wigner.

\end{document}